# RAPID 3D RECONSTRUCTION OF INDOOR ENVIRONMENTS TO GENERATE VIRTUAL REALITY SERIOUS GAMES SCENARIOS

*Zhenan Feng & Vicente A. Gonzalez*
The University of Auckland, Department of Civil and Environmental Engineering, New Zealand

*Ling Ma & Mustafa M.A. Al-Adhami*
The University of Huddersfield, Department of Architecture, UK

*Claudio Mourgues*
Pontificia Universidad Catolica de Chile, Department of Engineering and Construction Management, Chile

**ABSTRACT:** Virtual Reality (VR) for Serious Games (SGs) is attracting increasing attention for training applications due to its potential to provide significantly enhanced learning to users. Some examples of the application of VR for SGs are complex training evacuation problems such as indoor earthquake evacuation or fire evacuation. The indoor 3D geometry of existing buildings can largely influence evacuees' behaviour, being instrumental in the design of VR SGs storylines and simulation scenarios. The VR scenarios of existing buildings can be generated from drawings and models. However, these data may not reflect the 'as-is' state of the indoor environment and may not be suitable to reflect dynamic changes of the system (e.g. Earthquakes), resulting in excessive development efforts to design credible and meaningful user experience.

This paper explores several workflows for the rapid and effective reconstruction of 3D indoor environments of existing buildings that are suitable for earthquake simulations. These workflows start from Building Information Modelling (BIM), laser scanning and 360-degree panoramas. We evaluated the feasibility and efficiency of different approaches by using an earthquake-based case study developed for VR SGs.

## 1. INTRODUCTION

Virtual Reality (VR)-based Serious Games (SGs) are becoming popular in recent years (Connolly et al. 2012). VR SGs have been applied widely for training and education purposes in different domains (Feng et al. 2018). One of the significant applications of training is emergency evacuation training such as fire evacuation (Duarte et al. 2014) and indoor earthquake evacuation (Lovreglio et al. 2018). These evacuation trainings aim to train building occupants in a virtual environment and equip them with the best evacuation practice, so that they can correctly behave and respond when facing such emergencies in their daily use buildings. In that respect, indoor 3D environments of existing buildings are instrumental in the design of VR SGs storylines and simulation scenarios, which can largely influence building occupants' behavior by providing a similar evacuation experience. Indoor 3D environments can be reconstructed from drawings and models. However, the modelling process may be time-consuming. Moreover, the generated data may not reflect the "as-is" state of the indoor environment and may not be suitable to reflect dynamic changes of the system (e.g., earthquakes), resulting in excessive development efforts to design credible and meaningful user experience. Therefore, there is a need to explore innovative approaches to overcome the limitations mentioned above. Such innovations should be able to rapidly reconstruct indoor 3D environments to become available in VR SGs. Therefore, this research examined and evaluated the effectiveness to rapidly create VR SG environments through three 3D reconstruction workflows, using Building Information Modelling (BIM), laser scanning, and 360-degree panoramas.

BIM can be defined as "a modelling technology and associated set of process to produce, communicate and analyze building models" (Eastman et al. 2011). BIM is able to produce building models enriched with massive amounts of information towards different analysis and management purposes. One of them is 3D visualizations, in which users can navigate freely to view each detail of the models. The models are composed of individual components representing different objects of buildings such as walls, doors, floors, and furniture. Resulting from that, objects in the models can be manipulated to make dynamic changes, thus, to generate VR SGs simulation. Ruppel and Schatz (2011) showed that BIM could be coupled with game engines to develop fire safety evacuation SGs. Loverglio et al. (2018) proposed a BIM-based VR SG for earthquake evacuation training.

Laser scanning is a surveying approach that can rapidly capture the "as-is" state of a facility with a laser scanner

(Xiong et al. 2013). The data obtained from laser scanning, which is a set of distance measurements and images of surfaces visible from the sensor's viewpoint, is represented as point clouds (Xiong et al. 2013). A laser scanner can be placed in various locations throughout a building so that the generated point clouds from each location can be clustered to obtain a complete point cloud model of building indoor environments. Thus, point cloud models have the potential to be applied as the gaming environment for VR SGs simulation representing the "as-is" state of building indoor environments. Bruno et al. (2010) illustrated a complete methodology to develop VR systems for cultural heritage using laser scanning data.

360-degree panorama captures an un-modeled view of the real environment (Pereira et al. 2017). 360-degree panoramas can be obtained by 360-degree panoramic cameras, providing unbroken views of the whole environment surrounding the cameras (Pereira et al. 2017). Then, 360-degree panoramas can be applied to develop a VR, which looks identical to the real world. Users can be immersed into such VR simulations given a "sense of presence, of being there" resulting from the realistic panoramic views (Bourke 2014). 360-degree panoramas have been applied in various VR applications. Pereira et al. (2017) developed an interactive panoramic scene of construction sites using panoramic photos and videos. Gheisari et al. (2016) showed that an augmented panorama could provide a location-independent VR experience.

The feasibility and efficiency of each technique for generating VR SGs scenarios are different due to the variety of original data formats and modelling processes. Each technique has different inherent abilities. For instance, BIM may facilitate a high level of dynamic changes while laser scanning may offer a high level of fidelity of virtual environments. Therefore, this study conducted a pilot case study to examine each workflow for the rapid and effective reconstruction of 3D indoor environments of existing buildings applied in earthquake simulations.

## 2. FROM BIM TO VR SG

Models generated by BIM consist of complete building components that make them suitable for virtual simulations (Bille et al. 2014). Such simulations can be developed by game engines. A game engine is a software development platform used for developing video games (Bille et al. 2014). One popular game engine is Unity (https://unity3d.com/). Unity is a cross-platform game engine whose primary purposes are developing 3D and 2D video games and simulations for different operation systems. Unity also has the capability of adding VR features to games and simulations, which makes it an ideal tool for VR SGs development. The general process from BIM to game engines to develop virtual simulations is illustrated by Bille et al. (2014) in Figure 1.

Fig. 1: General BIM to VR SGs process (Bille et al. 2014)

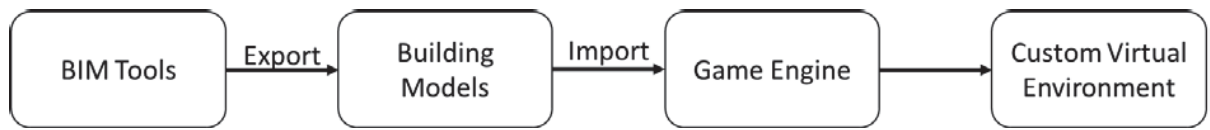

There are numbers of BIM tools available to generate building models. One of them is Autodesk Revit (www.autodesk.com), which is widely adopted by the architecture, engineering and construction (AEC) industry. Revit generates building models with parametric objects. These objects are individual with their own properties such as rendered materials, which offers the opportunity to set building models as the virtual environment for VR SGs.

Revit supports exporting models in various formats for use in other software. One fundamental format is FBX format, which stores information of polygonal mesh models. The FBX format contains geometric and material data and is directly readable by Unity and other 3D modelling software such as Autodesk 3ds Max (www.autodesk.com) or Blender (www.blender.org).

However, models exported from Revit in FBX format do not contain appropriate materials for representing textures in Unity. Revit exports FBX file with Autodesk materials while Unity can only read standard materials. Therefore, an additional process is required to convert Autodesk materials to standard materials (CIC Research Group 2016). Bille et al. (2014) suggested the conversion to be done by 3ds Max, given that there is a commercial script called Universal Material Converter (www.3dstudio.nl) which can directly convert Autodesk materials to standard materials. After the material conversion, the models are ready for developing VR SGs by game engines. The entire importation workflow is illustrated in Figure 2.

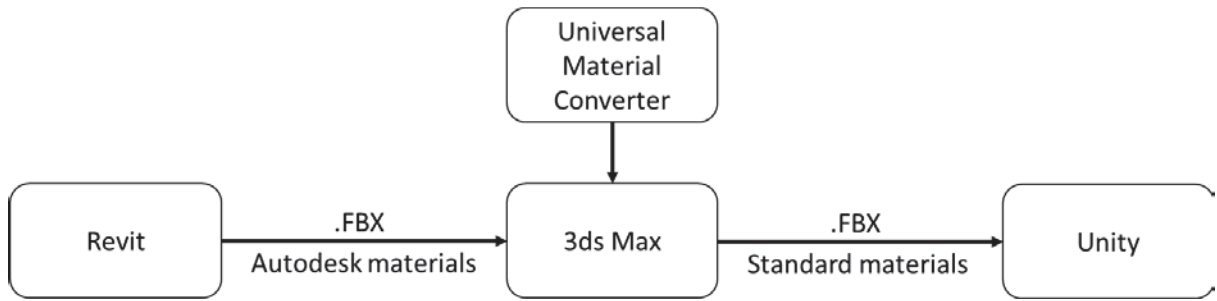

Fig. 2: Revit to Unity importation workflow (Bille et al. 2014)

Apart from that, there is also another free plugin called Walk-Through-3D (apps.autodesk.com) available for Revit, which can export 3D model geometry with standard materials from Revit to Unity directly.

Despite the straightforward importation workflow, BIM also has its limitations in terms of VR SGs development. One is that the models generated by BIM contain rich information and details which may be redundant for VR SGs, thus, reducing VR SGs development efficiency (Lovreglio et al. 2018). The cause of this issue is related to the essence of BIM as the native ability of BIM is for building design and management. For instance, a wall created by Revit includes multiple layers such as core structure, substrate, thermal layer, and finishing components. In addition, each layer has its own geometric data (e.g. thickness) and meta-data (e.g. material). Such rich information and details are not necessary for VR SGs development. A wall in a virtual environment can be simply represented as one single hexahedron with surface materials. However, the models converted from BIM with this rich information and details may become too heavy to be efficiently manipulated in game engines. Also, heavy models can reduce frame rate which may lead to an uncomfortable gaming experience (Johansson et al. 2014). Therefore, when developing VR SGs with a complex building model, acceleration techniques such as occlusion culling and hardware instancing need to be applied in game engines in order to improve performance (Johansson et al. 2014).

## 3. FROM LASER SCANNING TO VR SG

Laser scanning is adopted in the AEC industry for surveying, carried out by laser scanners. The initial output data of laser scanning is point clouds, which contained dense points with 3D measurements and images of the surface of a facility's "as-is" condition (Tang et al. 2010). Point clouds can represent indoor geometry with realistic colors of a building, which can be applied as gaming environment for VR SGs. In order to achieve that, a process is required to make point clouds data readable by game engines such as Unity.

One possible solution is to convert point clouds to mesh models in FBX format, which can be directly read by Unity. Lin et al. (2004) introduced an algorithm which can reconstruct a polygonal mesh from a given point cloud. The reconstructed mesh represents topologically correct surfaces of objects with numerous polygons. The initial polygonal mesh is in wireframe mode, which means only the points and lines are framed to describe the edges in a transparent drawing (Remondino 2003). To represent the images of surface in a facility, another process called texture mapping is required. Textures are mapped onto the polygon surface with the color at each pixel which is derived from the images (Remondino 2003). After texture mapping, the image-based mesh models converted from point clouds are ready for further applications such as game development. The general process from laser scanning to game engines to develop virtual simulations is illustrated in Figure 3.

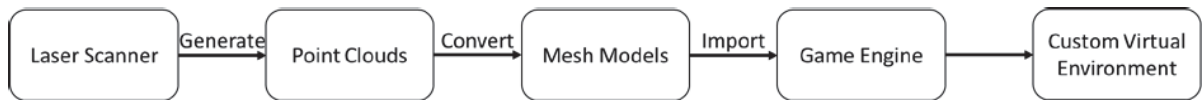

Fig. 3: General laser scanning to VR SGs process

Point clouds generated from laser scans can be stored in an E57 format, which is designed to be an open standard for storing data produced by 3D imaging systems (Huber 2011). In order to convert point clouds to polygonal mesh models, one toolset called Autodesk ReCap Pro (www.autodesk.com) can be adopted, which is a "Reality Capture" program for working with native point clouds from laser scans. ReCap Pro includes two programs, one is ReCap, and another is ReCap photo. ReCap can read original point clouds data and generate image-based mesh models on Autodesk cloud servers. ReCap photo can read and edit the generated mesh models and save them in FBX format. Figure 4 illustrates the entire workflow.

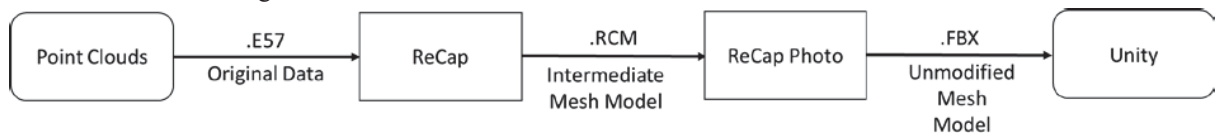

Fig. 4: Point clouds to Unity importation workflow

The converted mesh model is a single undivided mesh connecting all the objects captured by laser scanning. However, this is one of its limitations for VR SGs development. With a single mesh representing all the visible surfaces of objects, the capacity to manipulate each object in order to make dynamic changes is limited. One possible solution is to edit the entire single mesh in 3D modelling software in order to divide objects within the mesh. After that, the divided objects can be manipulated to perform dynamic changes. However, the main following issue is that once the divided objects being moved away from their original positions, there will be hollows left on the rest entire mesh since those places are invisible to laser scanners, thus, leaving no point clouds captured by laser scanning. Another possible solution is to provide an augmented virtual experience by adding additional virtual objects in game engines to perform dynamic changes, while the main mesh remains static as background environment.

Another limitation is that mesh models may contain large numbers of polygons (i.e., over millions of polygons), which can make computers inefficient to process, thus, reducing the frame rate and leading to an uncomfortable gaming experience (Papageorgiou, Platis 2015). In order to optimize performance, one possible solution is to reduce polygons in 3D modelling software with mesh simplification functions (Papageorgiou, Platis 2015). Another supplementary solution is to use acceleration techniques such as occlusion culling and hardware instancing as we suggested in the previous BIM workflow (Johansson et al. 2014).

## 4. FROM 360-DEGREE PANORAMAS TO VR SG

360-degree panoramas can provide immersive visualizations of the real world (Gheisari et al. 2016). 360-degree panoramas can be created by stitching multiple photos taken from a rotating robotic camera (Peleg, Ben-Ezra 1999). With current new technologies, 360-degree panoramas can also be generated by panoramic cameras with fish-eye lens in few seconds without manual process. Panoramic cameras capture a complete scene as a single image that can be viewed by rotating about a single central position (Wikipedia 2018). 360-degree panoramas are ideal to be applied for VR. The general process from 360-degree panoramas to game engines to develop virtual simulations is illustrated in Figure 5.

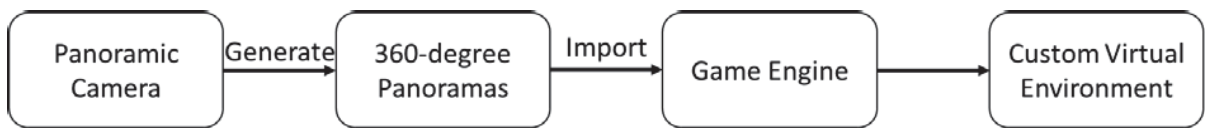

Fig. 5: General 360-degree panoramas to VR SGs process

360-degree panoramas taken by panoramic cameras are usually in JPG format which can be directly read by game engines such as Unity. There is no additional process required to convert the original data. Figure 6 illustrates the importation workflow.

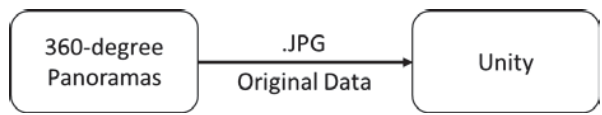

Fig. 6: 360-degree panoramas to Unity importation workflow

However, 360-degree panoramas also have limitations to be extended to VR SGs. One is that panoramas are only planar images without depth so that participants are not able to navigate "inside" the images. Another is that objects are captured as static 2D pictures in panoramas which make them impossible to be manipulated in game engines to make dynamic changes.

In order to overcome the navigation limitation, one possible solution is to take a series of panoramas on the potential navigation path and teleport users from one point to another (Pereira et al. 2017). One famous application of this solution is Google Street View (map.google.com), which teleports users to move forward and backward. Sequential panoramas on the navigation path can be connected by scripts in game engines to implement teleportation.

Regarding the limitation of dynamic changes, one possible solution is to add additional layers of information on static panoramas (Gheisari et al. 2016). In other words, objects in panoramas remain in an "as-is" state while virtual objects and information can be added to augment virtual experiences. Such augmented information can be created and manipulated in game engines. As a result, users can still receive a perception of simulated scenarios which are delivered by augmented information.

## 5. EVALUATION OF 3D RECONSTRUCTION APPROACHES USING A PILOT STUDY

This research aims to compare and evaluate the different 3D reconstruction workflows mentioned above towards the rapid development for VR SGs. The feasibility and efficiency of different workflows were examined by using an earthquake-based pilot case study developed for VR SGs.

### 5.1 Research subject

An area of the Engineering VR/AR Lab of the University of Auckland was selected as the reconstruction object. The floor area is approximately 6m x 5m, and the clearance height is approximately 2.6m. The lab includes various objects such as desk, chairs, computers, drawers, partition walls, and cardboard boxes. The photo showing the actual state of the lab is illustrated in Figure 7(a).

### 5.2 Apparatus

Three different reconstruction workflows, namely BIM, laser scanning, and 360-degree panoramas, were adopted to reconstruct the 3D indoor environment of the Engineering VR/AR lab.

The PC we used to run the VR SGs earthquake scenarios is equipped with 16GB of RAM, a Nvidia GTX 980m graphic card with 8GB of memory, and a 2.60GHz Intel i7-6700HQ processor. The VR headset we used is Oculus Rift (www.oculus.com). Oculus Rift is a consumer-grade VR headset with 1080x1200 resolution per eye, and a 110-degree field of view.

The BIM workflow was carried out on the same PC as mentioned above. Cameras and measure tapes were adopted to run a simple survey as no plan drawings were available.

The laser scanner we used is FARO Focus3D X 330 (www.faro.com). The scanner can scan objects up to 330 meters away and in direct sunlight, providing precise 3D models in a photo-realistic style.

The 360-degree panoramic camera we used is Nikon KeyMission 360 (www.nikon.co.nz). This camera is equipped with two fish-eye lenses that can directly capture panoramas with no manipulation from users. The camera captures a complete 360-degree panorama with a single shot. The output resolution is 4k.

### 5.3 Research design

Each workflow was applied individually to capture data, reconstruct 3D environment, and develop earthquake scenarios for VR SGs. As a result, a total of three earthquake scenarios were developed based on the virtual environments generated from each reconstruction workflow. The detailed development process will be described in section 5.4.

Tang et al. (2010) proposed performance measures for modelling as-built BIMs, which can be catogrized into three aspects: 1) measures of algorithm design; 2) measures of environmental conditions; 3) measures of modelling performance. Inspired by these measures, we formulated six performance dimensions to evaluate the feasibility and efficiency of each workflow. Table 1 shows the detailed description and assessment criteria of each performance dimension. Each workflow in each performance dimension was compared with each other and ranked from low to high level with a score ranged from 1 to 3, where 1 stands for low ranking, 3 stands for high ranking. The score is a relative value which is only to show different rankings of each workflow in each performance dimension.

Table 1: Performance dimensions of different 3D reconstruction workflows

| Performance Dimension | Description | Score Criteria | Measurement Type |
| --- | --- | --- | --- |
| **Level of fidelity** (static environment, before earthquakes) | How close is the static virtual environment to the "as-is" state of the real world? | 3 stands for the highest fidelity | Subjective assessment with qualitative measure |
| **Level of fidelity** (dynamic environment, during earthquakes) | How realistic is the earthquake simulation? | 3 stands for the highest fidelity | Subjective assessment with qualitative measure |
| **Level of dynamic capability** | To what extend can dynamic changes be performed in the virtual environment? | 3 stands for the highest dynamic capability | Subjective assessment with qualitative measure. |
| **Time requirement** | How much time is consumed for completing the entire approach? | 3 stands for the least time consumption | Subjective assessment with quantitative measure |
| **Model complexity** | How many polygons does the virtual environment have? | 3 stands for the lowest polygons | Objective assessment with quantitative measure |
| **Computational complexity** | How many frame rates per second (FPS) does the virtual environment achieve when running earthquake simulation in VR headset? | 3 stands for the highest FPS | Objective assessment with quantitative measure |

The first two performance dimensions namely level of fidelity (static) and level of fidelity (dynamic) were examined subjectively by five undergraduate students. Firstly, each student was asked to experience three static environments (before earthquakes) in order to compare them with the real world to rank the level of fidelity (static) of each workflow, where the highest fidelity was given 3 score and the lowest got 1 score. As a result, mean values of each workflow were calculated and used to rank each workflow accordingly. Secondly, each student was asked to experience three dynamic environments (during earthquakes) in order to rank the level of fidelity (dynamic) of each workflow, where the highest fidelity was given 3 score and the lowest got 1. As before, each workflow was ranked based on the mean values.

Level of dynamic capability and time requirement were examined subjectively by the researcher who developed these three earthquake scenarios. Each workflow was evaluated and ranked with the capability to perform dynamic changes. As for time requirement, the time consumed for developing the complete VR scenarios by each workflow was recorded and used as indicators to rank each workflow.

The last two performance dimensions namely model complexity and computational complexity were examined objectively by comparing the performance data recorded by the computer running the virtual scenarios. The total number of polygons of each virtual environment was used as indicators of model complexity, while the frame rates per second (FPS) of each virtual environment running earthquake simulation in VR headset was used as indicators of computational complexity.

### 5.4 Procedure

A BIM model of the lab including furniture was developed by a researcher skilled in Revit. However, there were no plan drawings available so that the geometry information was obtained from a simple field survey. Then the 3D model was imported from Revit into Unity by the plugin called Walk-Through-3D. Figure 7(b) shows the BIM-based virtual environment in Unity. After that, the earthquake simulation was developed in Unity by using C# script enabling physics engine to shake and bring down objects in the lab. A cloud of dust (virtual particle system) was also added to the simulation. The VR function was enabled supporting Oculus Rift in Unity. Figure 7(c) shows the BIM-based post-earthquake environment.

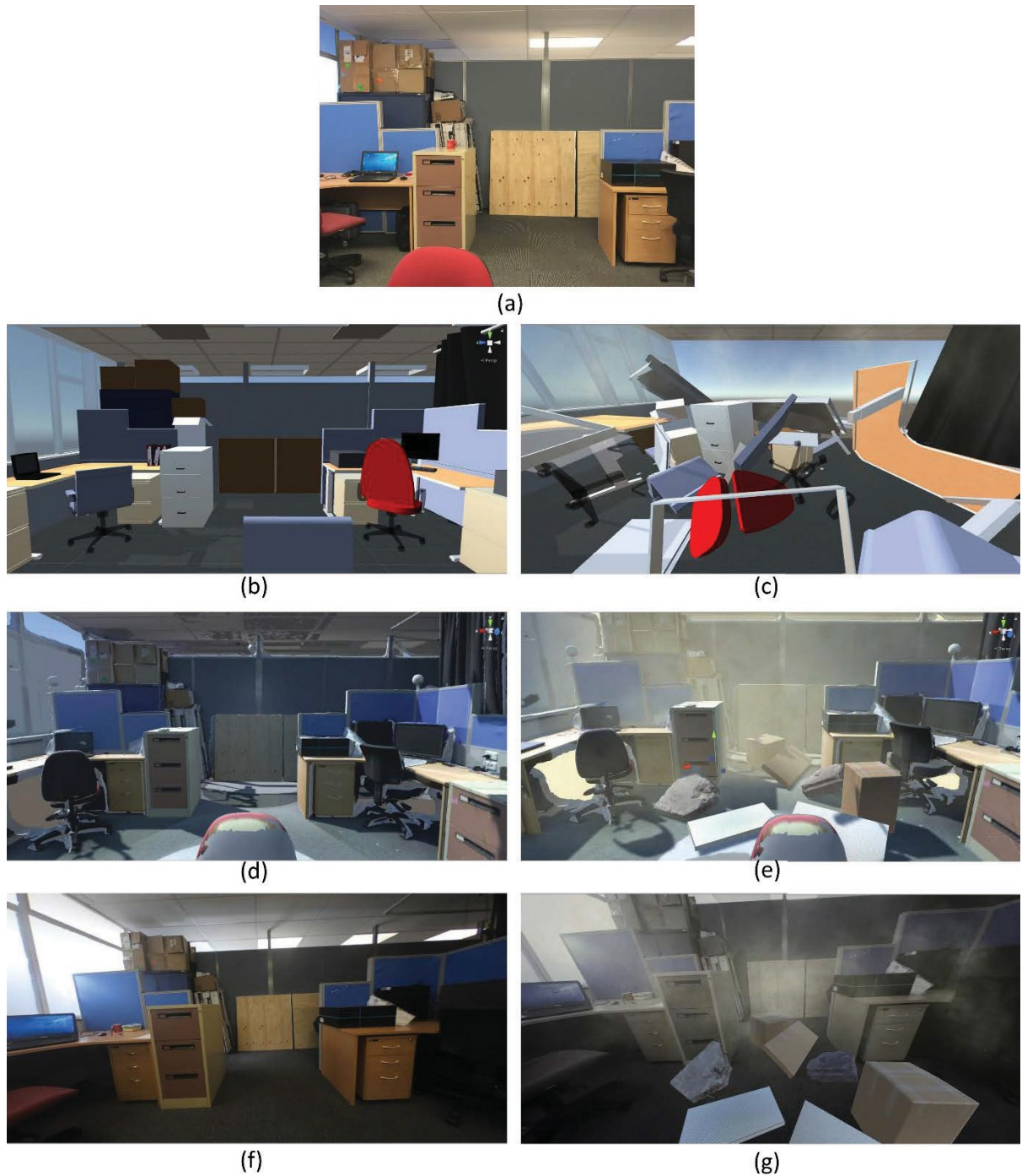

Fig. 7: (a) Actual state of the lab, (b) BIM-based 3D environment, (c) BIM-based post-earthquake environment, (d) Laser scanning-based 3D environment, (e) Laser scanning-based post-earthquake environment, (f) 360-degree panorama-based 3D environment, (g) 360-degree panorama-based post-earthquake environment

Two laser scans were produced by the laser scanner in two positions in the lab in order to get maximum coverage of the environment. Two scans were registered and clustered in Faro Scene and exported as point clouds in E57 file format. ReCap was used to convert point clouds to a polygonal mesh model in FBX format. Then the mesh model was imported to Unity. Figure 7(d) shows the laser scanning-based virtual environment in Unity. After that, additional layers of information were added in Unity to augment the earthquake simulation including falling objects (virtual 3D objects with materials) and a cloud of dust (virtual particle system). The main camera was shaken in Unity by using C# script in order to provide a perception of shaking experience as earthquakes. The VR function was enabled supporting Oculus Rift in Unity. Figure 7(e) shows the laser scanning-based post-earthquake environment.

One 360-degree panorama was taken by the panoramic camera. The panorama was directly imported to Unity. Figure 7(f) shows the 360-degree panorama-based virtual environment in Unity. Additional layers of information were added in Unity to augment the earthquake simulation including falling objects (virtual cubes with materials), flickering lights (change brightness of panorama), and a cloud of dust (virtual particle system). The main camera was shaken in Unity by using C# script in order to provide a perception of shaking experience as earthquakes. The VR function was enabled supporting Oculus Rift in Unity. Figure 7(g) shows the 360-degree panorama-based post-earthquake environment.

## 5.5 Results

Three earthquake simulations were developed for VR SGs based on three different virtual environments. Mean values of rankings for levels of fidelity are shown in Table 2. Time consumed to complete each workflow was recorded in Table 3. Polygons and FPS of each simulation at runtime, where the former stands for model complexity and the latter stands for computational complexity, were obtained from statistics provided by Unity and recorded in Table 4. Based on these results, each workflow was ranked in performance dimensions listed in Table 1, which are shown in Table 5.

Table 2 Mean values of rankings for each workflow in terms of level of fidelity

|  | BIM | Laser Scanning | 360-Degree Panorama |
|---|---|---|---|
| Level of fidelity (static) | 1.6 | 2.6 | 1.8 |
| Level of fidelity (dynamic) | 2.6 | 1.8 | 1.6 |

Table 3 Time consumed to complete each workflow

|  | BIM | Laser Scanning | 360-Degree Panorama |
|---|---|---|---|
| Survey time (optional) | 0.5 hour | 0 | 0 |
| Model creation time | 7 hours | 1 hour | 0.1 hour |
| Model treatment time (conversion) | 1 hour | 2 hours | 0 |
| Simulation development time | 1.5 hours | 1.5 hours | 1.5 hours |
| Total time | 10 hours | 4.5 hours | 1.6 hours |

Table 4 Model and computational complexity of each simulation

|  | BIM | Laser Scanning | 360-Degree Panorama |
|---|---|---|---|
| Polygons | 2.1 million | 13.6 million | 35.5 thousand |
| FPS | 46.1 | 42.3 | 86.3 |

Table 5 Performance rankings of each workflow

| Performance Dimension | BIM | Laser Scanning | 360-Degree Panorama |
|---|---|---|---|
| Level of fidelity (static) | 1 | 3 | 2 |
| Level of fidelity (dynamic) | 3 | 2 | 1 |
| Level of dynamic capability | 3 | 2 | 1 |
| Time requirement | 1 | 2 | 3 |
| Model complexity | 2 | 1 | 3 |
| Computational complexity | 2 | 1 | 3 |

# 6. CONCLUSIONS AND DISCUSSIONS

This research examined three 3D reconstruction workflows for indoor environments suited to VR SGs. The results indicated that each workflow has its unique advantages and limitations.

BIM is a comprehensive approach to reconstruct indoor environments due to the nature of its modelling ability and process. This workflow can be time-consuming if it is for a large building with large amounts of objects. Besides, BIM may not represent the "as-is" state of buildings if the modelling is based on old drawings. When modelling, it is also difficult to get the objects similar to the ones in the real world. In this research, we spent hours to get the objects as close as possible to the real ones. Even though, the reconstructed model still misses some details which make it different to the "as-is" state of the indoor environment. Apart from that, the heavy model with rich information can lead to a low development efficiency of VR SGs in game engines and a low runtime performance. However, BIM also shows the advantage in terms of dynamic ability. BIM is able to perform a high level of dynamic changes given that all the objects in models are individual, which offers BIM high extensibility to game engines for further manipulation and development (Bille et al. 2014).

Laser scanning is a faster 3D reconstruction approach compared to BIM. It also provides a higher level of fidelity which can reflect the "as-is" state of the indoor environment. Even though the objects are not able to be manipulated in an individual basis, laser scanning-based virtual environment still shows a certain capability to be extended to VR SGs by enhancing dynamic changes through additional virtual objects as augmentation. However, the major limitation is that the mesh model converted from point clouds contain massive numbers of polygons, which largely depends on the complexity and regulation of original geometry. This issue can lead to a relatively low runtime performance. An additional process is required to reduce polygon numbers before importing mesh model into Unity.

Regarding 360-degree panorama, it is the fastest approach with the least computational complexity. And it is interesting to find that even though 360-degree panorama provides an un-modeled view of surroundings, it gives a weaker perception in terms of level of fidelity of static environment. Besides, the major limitation is that limited dynamic changes and interaction can be made for VR SGs. In this research, 360-degree panoramas showed limited ability to be extended to earthquake scenarios due to only few earthquake damages can be represented and such dynamic representation has a relatively low level of fidelity. However, 360-degree panoramas may have better performance in other fields that require less dynamic changes such as fire simulation (can be augmented by flame and smoke mainly) or construction management (Pereira et al. 2017).

This research also has some limitations. One is that the experiment was limited to a situation where there was no navigation through the virtual environments. Each workflow may require different navigation solutions, which can largely influence user experience. Another limitation is that due to different data formats and modelling processes, each earthquake scenarios was generated using different algorithms and visual effects, which leads to the comparison of each workflow is not rigorous.

This research provides an insight of each workflow based on the evaluation of feasibility and efficiency. It is difficult to draw a conclusion that which workflow is the best towards VR SGs. Selection criteria and actual performance depend on various aspects. For instance, for large and complex buildings, time consumption of BIM can increase exponentially due to models need to be created manually, while computational complexity of laser scanning can increase exponentially due to numerous of polygons for complex geometry. Besides, different algorithms and visual and sound effects applied for VR SGs can also influence the performance. Possible future research could be to establish a framework mixing multiple workflows in order to take the advantages of each approach, for instance, using laser scanning and photogrammetry to create an "as-is" BIM model for VR SGs development.